\documentclass[epj-spec]{svjour}
\usepackage{graphics}
\def\la{\langle}
\def\ra{\rangle}
\def\beq{\begin{equation}}
\def\eeq{\end{equation}}
\def\be{\begin{eqnarray}}
\def\ee{\end{eqnarray}}

\def\k2av{\la k_T^2\ra}
\newcommand{\f}[2]{\frac{#1}{#2}}
\newcommand{\dd}{ {\textrm d}}
\begin{document}
\title{Pion production in dAu collisions at RHIC energy}
\author{P. L\'evai\inst{1,2}\fnmsep\thanks{\email{plevai@rmki.kfki.hu}}   
\and G. Papp\inst{3}
\and G.G. Barnaf\"oldi\inst{1,4} 
\and G. Fai\inst{4} }
\institute{RMKI 
Research Institute for Particle and Nuclear Physics,
      PO Box 49, Budapest, 1525, Hungary
\and 
      Department of Physics, Columbia University, 
      538 West 120th Street, New York, NY 10027, USA
\and 
      Department for Theoretical Physics,
      E{\"o}tv{\"o}s University, 
      P{\'a}zm{\'a}ny P. 1/A, Budapest 1117, Hungary 
\and 
      Center for Nuclear Research, Department of Physics,  
      Kent State University, Kent, OH 44242
}
%
%
\abstract{
We present our results on neutral pion ($\pi^0$) production
in $pp$ and $dAu$ collisions at RHIC energy.  Pion spectra
are calculated in a next-to-leading
order (NLO) perturbative QCD-based model.
The model includes the transverse component of the
initial parton distribution (``intrinsic $k_T$'').
We compare our results to the available experimental data from RHIC,
and fit the data with high precision.
The calculation tuned this way is repeated for the $dAu$ collision,
and used to investigate the interplay of shadowing and
multiple scattering at RHIC.
The centrality dependence of the nuclear modification
factor shows a measurable difference between different
shadowing parameterizations.
} 
\maketitle

\section{Prologue}

With this paper we would like to pay tribute to J\'ozsef Zim\'anyi
(Jozs\'o) who passed away in September 2006. Jozs\'o was an  
enthusiastic supporter of the Hungarian participation in the relativistic heavy 
ion program at the Relativistic Heavy Ion Collider (RHIC) at Brookhaven National 
Laboratory. He had a detailed vision about quark-gluon
plasma formation at RHIC, and he was eager to confront 
the real world of experimental data with his ideas. In the last decade of 
the construction phase of RHIC he predicted many observables,
including hadron numbers and ratios~\cite{LastCall}. He realized 
that the era of high-$p_T$ physics and QCD is coming to nuclear physics,
because the RHIC detectors extended the measurable transverse momentum region 
much beyond the SPS values of $p_T=3-4$ GeV/c. He followed the 
progress in this subfield and was always happy to discuss related questions.

Jozs\'o joined the Hungarian experimental PHENIX group from its foundation and
supported continuously the Hungarian activity by full heart and by his authority.
The first PHENIX paper on which his name appears is entitled
{\it "Absence of suppression in particle production at large transverse
momentum in $\sqrt{s_{NN}}$= 200 GeV in $d+Au$ collisions"}~\cite{PHENIXdAu},
and has become a fundamental reference in the quest for understanding hadron 
suppression and jet energy loss in deconfined matter in $AuAu$ collisions.
Jozs\'o is among the authors of many other $dAu$ papers,
including a detailed analysis of centrality dependence of the
nuclear modification factor in $dAu$ collisions~\cite{PHENIXdAu07centr},
the results of which will be used in our present theoretical analysis.
The results of Ref.~\cite{PHENIXdAu06} about 
identified particle production and Ref.~\cite{PHENIXdAu07} about charge 
particle production will also be commented on.

Our group has performed theoretical investigations connected to the 
$pp$, $dAu$, and $AuAu$ collisions at RHIC energies 
in the framework of a RMKI-ELTE-Kent-Columbia collaboration.
We would like to posthumously thank Jozs\'o with this paper
on $dAu$ collisions for his continuous interest in our work
and for his support to keep our theoretical collaboration vital
and successful in high-energy nuclear collision research.

\section{ Introduction}

Recent experimental data on mid-rapidity $\pi^0$ production 
in the transverse momentum region 2 GeV/c $< p_T <$ 15 GeV/c
measured at the Relativistic Heavy Ion Collider (RHIC)
in $AuAu$ collisions at $\sqrt{s}=$ 130 AGeV and 200 AGeV
display surprising evidence
of new phenomena appearing in hot dense matter.
Here we focus on the property that
the $\pi^0$ transverse momentum spectrum
shows a strong suppression~\cite{PHENpi0130,PHENpi0200,STARsup} 
in central collisions
compared to theoretical expectations based on binary collision dynamics 
in perturbative quantum chromodynamics (pQCD) calculations
(see e.g. Refs.~\cite{Wang01,Papp02,YZ02}). A
similar suppression pattern was observed for charged 
hadrons~\cite{STARh200}.  Introducing final state
interactions, especially jet energy 
loss~\cite{gptw,mgxw92,GLV,BDMS02,Zakharov,Wiedemann,GLV02},
this suppression pattern can be reproduced  both  
at 130 AGeV~\cite{Lev02,XNW02s130} and 200 AGeV~\cite{XNW03s200,BLPF03}.
The entanglement 
of initial and final state interactions in $AuAu$ collisions
and the complexity of the theoretical calculation generated an 
increased interest in simpler reactions, 
such as the $dAu$ collision. In this case final state
interactions play a minor role, and the initial state interactions
(initial multiscattering, shadowing) can be investigated cleanly.
The obtained $dAu$ data from 
RHIC~\cite{RHIC_user03_PHENIX,RHIC_user03_PHOBOS,RHIC_user03_STAR}
stimulated further interest
in theoretical calculations of pion production in this reaction.

In a pQCD-based leading order (LO)
parton model a systematic analysis of existing $pA$ data 
has been completed~\cite{YZ02}  and our calculation will be based
on results obtained there. A LO calculation of $dAu$ 
was reported in Ref.~\cite{Vitevlo}. 
Here we perform a $dAu$ computation in next-to-leading order (NLO), 
utilizing recent developments for $pp$ collisions~\cite{pgNLO}.
Successful calculations based on the physics of gluon saturation
have also been carried out for this system~\cite{CGC1,CGC2}.

In this paper we take advantage of the availability of detailed 
(impact-parameter selected) data and calculate the impact-parameter
dependence of nuclear modification factor $R_{dAu}$ in our pQCD
based model. 

The paper is organized as follows. In Section 2 we review the NLO 
pQCD-improved parton model augmented with the intrinsic transverse momentum
distribution for $pp$ collisions.
We compare calculations to $pp$ results at midrapidity
at $\sqrt{s}=200$ GeV  and use these data to tune the parameters of
the model.
In Section 3, we discuss calculational details for the
$d+Au \rightarrow \pi^0 + X$ reaction at $y=0$ in the transverse 
momentum region $p_T> 2$ GeV/c. 
In Section 4 the $R_{dAu}$ nuclear modification factor is extracted from
the data, and we investigate its impact parameter dependence.
In particular, we study simple shadowing prescriptions without 
and with impact parameter dependence.
In Section 5, we discuss our results.

\section{The pQCD improved parton model with \mbox{intrinsic $k_T$}}

The invariant cross section for neutral pion production in a
$pp$ collision can be described in the NLO pQCD-improved parton model 
on the basis of the factorization
theorem as a convolution~\cite{pgNLO,Aversa89,Aur00}:
\begin{eqnarray}
\label{hadX}
 E_{\pi}\f{\dd \sigma^{pp}}{\dd ^3p_\pi} &=&
        \f{1}{S} \sum_{abc}
  \int^{1-(1-V)/z_c}_{VW/z_c} \f{\dd v}{v(1-v)} \ 
  \int^{1}_{VW/vz_c} \f{ \dd w}{w} 
  \int^1 {\dd z_c} \nonumber \\
  && \ \int {\dd^2 {\bf k}_{Ta}} \ \int {\dd^2 {\bf k}_{Tb}}
        \, \, f_{a/p}(x_a,{\bf k}_{Ta},Q^2)
        \, f_{b/p}(x_b,{\bf k}_{Tb},Q^2) \cdot
   \nonumber \\
&& \cdot  
 \left[
 \f{\dd {\widetilde \sigma}}{\dd v} \delta (1-w)\, + \,
 \f{\alpha_s(Q_R)}{ \pi}  K_{ab,c}(s,v,w,Q,Q_R,Q_F) \right] 
 \f{D_{c}^{\pi} (z_c, Q_F^2)}{\pi z_c^2}  \,\,   ,
\end{eqnarray}
where we use a product approximation for the 
parton distribution functions (PDFs), 
\be
f(x,{\bf k}_{T},Q^2) = f(x,Q^2) g({\bf k}_{T}) \ .
\ee
Here, the function $f(x,Q^2)$ represents the standard NLO PDF  
as a function of momentum fraction $x$ at factorization scale $Q$,
$\dd {\widetilde \sigma}/ \dd v$ is the  Born cross section of the
partonic subprocess $ab \to cd$,
$K_{ab,c}(s,v,w,Q,Q_R,Q_F)$ is the corresponding higher order correction term,
and the fragmentation function (FF), $D_{c}^{\pi}(z_c, Q_F^2)$,
gives the probability for parton $c$ to fragment into a pion
with momentum fraction $z_c$ at fragmentation scale $Q_F$.
We use the conventional proton level ($S,V,W$) and parton level ($s,v,w$)
kinematical variables of NLO calculations
(see details in Refs.~\cite{pgNLO,Aversa89,Aur00}).
In our present study we consider fixed scales:
the factorization and the renormalization
scales are connected to the momentum of the intermediate jet,
$Q=Q_R=\kappa\cdot p_q$ (where $p_q=p_T/z_c$),
while the fragmentation scale is connected to the final hadron momentum,
$Q_F=\kappa \cdot p_T$.
The value of $\kappa$ can be varied in a wide range, 
$\kappa \ {\cal{2}} \ [0.3,3]$.

Our NLO calculation includes
the initial transverse-momentum distribution $g({\bf k}_T)$ of partons
(``intrinsic $k_T$''),  along the lines of 
Refs.~\cite{Wang01,YZ02,pgNLO,Wong98}. We demonstrated the
success of such a treatment at LO level in Ref.~\cite{YZ02}.
In our phenomenological approach the transverse-momentum distribution
is described by a Gaussian,
\beq
\label{kTgauss}
g({\bf k}_T) \ = \f{1}{\pi \la k^2_T \ra}
        e^{-{k^2_T}/{\la k^2_T \ra}}    \,\,\, .
\eeq
Here, $\langle k_T^2 \rangle$ is the 2-dimensional width of the $k_T$
distribution and it is related to the magnitude of the
average transverse momentum of a parton
as $\langle k_T^2 \rangle = 4 \langle k_T \rangle^2 /\pi$.

In our LO and NLO
investigations we use GRV LO PDFs~\cite{GRVLO} and
the MRST(cg) NLO PDFs~\cite{MRST01} for parton distribution functions. 
For fragmentation functions we use the 
KKP parameterization~\cite{KKP}, which has LO and NLO versions.
An advantage of the GRV and the MRST sets is that
they can be used down to very small scales ($Q^2\approx 0.25$ GeV$^2$/c$^2$).
Thus, they provide reasonable calculations at relatively
small transverse momenta, $p_T \geq 2$ GeV/c at our fix scales.

\begin{figure}
\hspace{2.5truecm}
\resizebox{0.66\columnwidth}{!}{  
\includegraphics{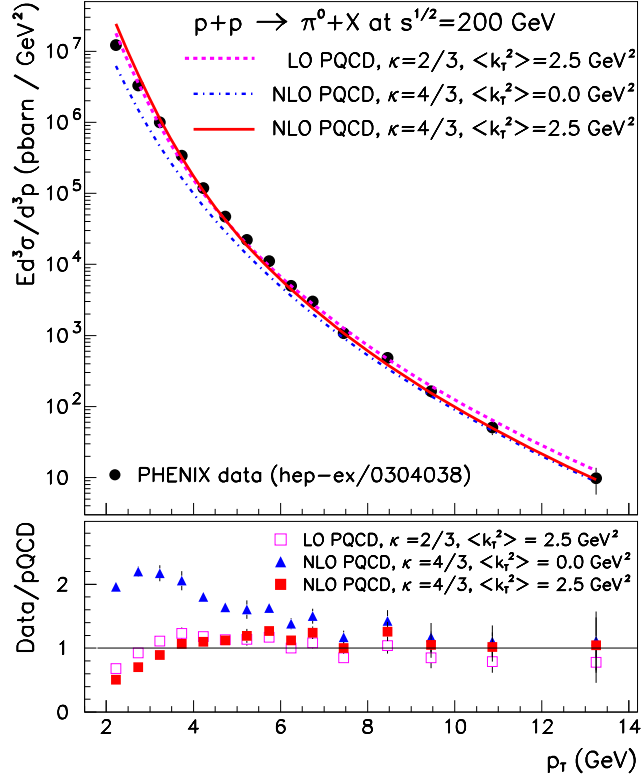}    
}   
\caption{ Invariant cross section   
of pion production in $p+p \rightarrow \pi^0 +X$ at $\sqrt{s}=200$ GeV.  
Upper panel displays spectra: LO result at scales $Q=(2/3)p_T/z_c,
\ Q_F = (2/3) p_T$ and $\langle k_T^2 \rangle=2.5$ GeV$^2$/c$^2$ (dashed line),
NLO results at scales  $Q=Q_R=(4/3)p_T/z_c, \ Q_F = (4/3) p_T$
without intrinsic $k_T$ (dash-dotted line) and
with $\langle k_T^2 \rangle=2.5$ GeV$^2$/c$^2$ (solid line).
Lower panel shows the data/pQCD ratios for LO (open squares),
NLO without (triangles) and
with (solid squares) intrinsic $k_T$ at the above parameter values.}
\label{fig:1}       
\end{figure}

Figure 1 displays our LO and NLO pQCD results at $\sqrt{s}=200$ GeV
for $pp\rightarrow \pi^0X$~\cite{PHENpi0200pp}
at different scales.
In the LO case (dashed line) we used $\kappa=2/3$ for both 
the factorization and the fragmentation scales to reproduce the data
at high $p_T$ within error bar. 
(This scale value is  
higher than $\kappa=1/2$ used in Ref.~\cite{YZ02} 
due to a readjustment triggered by the availability of 
precise data at high $p_T$ on $\pi^0$ production 
in place of the earlier UA1 data on charged hadrons.) 
To minimize the difference
between RHIC data and the model at low $p_T$ a  
$\langle k_T^2 \rangle = 2.5$ GeV$^2$/c$^2$ is included.
We obtained quite good agreement in LO (see open squares in the 
data/pQCD ratio, lower panel).

In the NLO calculation we use the scales $Q=Q_R=(4/3) p_q$ and  
$Q_F= (4/3) p_T$ to reproduce the central value of the experimental data
at high $p_T$ (dash-dotted line in the upper panel).
However, in the window  2 GeV/c $< p_T <$ 4 GeV/c this choice
underestimates the data by a factor of 2 (see filled triangles
in the lower panel). These calculational results are in agreement with 
the calculation reported in Ref.~\cite{PHENpi0200pp}, albeit 
with a different set of scales~\cite{note}.  
Since  we plan to
investigate multiscattering and the Cronin effect in $dAu$
collisions, which are most pronounced 
in the above momentum window, we consider an NLO parameter
set with scales $Q=Q_R=(4/3) p_q$ and  
$Q_F= (4/3) p_T$, 
together with an intrinsic $k_T$ value $\langle k_T^2 \rangle = 2.5$ GeV$^2$/c$^2$.

Fig.~1 indicates that we obtain good agreement both at high and low $p_T$  
with this choice (full line in the upper panel, filled squares in the lower
panel).
One can see that the results with this NLO parameter set and our
LO results are close to each other, even though
the scales differ by a factor 2. This is due to having the same 
width of the intrinsic $k_T$ distribution.  
The physically realistic nature of the value found for the width 
is corroborated by results from di-jet production at lower energies:
jet-jet correlations yield a similar value for $\k2av$
at ISR energies~\cite{Angelis80}. While
recent measurements of jet-jet correlations in $pp$ collision
at RHIC energies~\cite{STARdijet,PHENIXdijet} 
may further clarify the properties of the transverse
component of the PDFs,
these parameterizations fixed in $pp$ collisions will 
provide a solid basis to investigate nuclear collisions.

The precision of the NLO calculation at lower $p_T$ could be improved
introducing non-Gaussian or $p_T$-dependent intrinsic $k_T$
(see e.g. Refs.~\cite{GLVelas,gawron03}).
In the present analysis we keep a constant value for the width of the
Gaussian transverse momentum distribution in $pp$ collisions.
This way our results can be compared to the experimental data
 on measurable initial 
 dijet transverse momentum in a simple way.
This analysis has been accomplished~\cite{LFP06}, using recent data
on di-hadron correlation~\cite{PHENIXimbal}. The measured and 
theoretically extracted momentum imbalance value agrees
very well with the one applied in our calculations.

\section{ The parton model for $dAu$ collisions}

Considering the $dAu$ collision, the hard pion production cross section
can be written as an integral over impact parameter $b$, where
the geometry of the collision is  described in 
the Glauber picture:
\beq
\label{dAuX}
  E_{\pi}\f{\dd \sigma_{\pi}^{dAu}}{ \dd ^3p} =
  \int \dd ^2b \, \dd ^2r \,\, t_d(r) \,\, t_{Au}(|{\bf b} - {\bf r}|) \cdot
  E_{\pi} \,    \f{\dd \sigma_{\pi}^{pp}(\k2av_{pAu},\k2av_{pd})}
{\dd ^3p}
\,\,\, ,
\eeq
where the proton-proton cross section on the right hand side represents
the cross section from eq. (\ref{hadX}), 
but with the 
broadened widths of the transverse-momentum distributions (\ref{kTgauss}),
as a consequence
of nuclear multiscattering (see eq. (\ref{ktbroadpA})).  
Here $t_{A}(b) = \int \dd z \, \rho_{A}(b,z)$
is the nuclear thickness function (in terms of the density distribution 
of the gold nucleus, $\rho_{Au}$),
normalized as $\int \dd ^2b \, t_{Au}(b) = A_{Au} = 197$. 
For the deuteron, one could use a superposition of a $pAu$ and a $nAu$ 
collision, or a distribution for the nucleons inside the deuteron.
For a first orientation we follow Ref.~\cite{Vitevlo} 
in this regard, and apply
a hard-sphere approximation for the deuteron with $A=2$ for
estimating the nuclear effects. Also, since $\pi^0$ production
is not sensitive to isospin, we continue to use the notation ``$pA$''
when talking about the interaction of any nucleon with a nucleus.

The initial state broadening of the incoming parton
distribution function is accounted for by an
increase in the width of 
gaussian parton transverse momentum distribution in eq. (\ref{kTgauss}):
\beq
\label{ktbroadpA}
\k2av_{pA} = \k2av_{pp} + C \cdot h_{pA}(b) \ .
\eeq
Here, $\k2av_{pp}$ is the width of the transverse momentum distribution
of partons in $pp$ collisions, 
$h_{pA}(b)$ describes the number of {\it effective}
nucleon-nucleon (NN) collisions at impact parameter $b$,
which impart an average transverse momentum squared $C$.
The effectivity function $h_{pA}(b)$ can be written in terms of the
number of collisions suffered by the incoming proton in the target
nucleus, $h_{pA}(b)= \nu_A(b)-1$. Here, $\nu_A(b) = \sigma_{NN} t_{A}(b)$, 
with $\sigma_{NN}$ being
the inelastic nucleon-nucleon cross section.
Our preliminary results on the determination of the factor of
$C$ and $\nu(b)$ in an analysis in NLO
confirm the findings of Ref.~\cite{YZ02},
where the systematic analysis of $pA$ reactions 
was performed in LO and the characteristics
of the Cronin effect were determined  at LO level.
Following Ref.~\cite{YZ02}, we 
assume that only a limited number of semi-hard collisions 
(with maximum  $\nu_A(b)_{max} = 4$) 
contributes to the broadening,
and the factor $C =$ 0.4 GeV$^2$/c$^2$. To give an indication of the 
dependence of the results on the value of $C$, we carried out a
sample calculation with $C=1$ GeV$^2$/c$^2$. 

At RHIC energies, parton broadening can be understood in terms of
parton-level collisions. 
Elastic or small-angle-inelastic parton-parton collisions become 
responsible for the modification in the transverse momentum distribution, 
and this extra broadening will be superimposed to the original 
$\k2av_{pp}$ value~\cite{Papp02,accardicern}. 
The above effectivity function, $h_{pA}(b)$, can be connected to the 
effective partonic collision length. In this framework formation 
time effects  may result in a saturation-like behavior. 
In the following, we apply eq.~(\ref{ktbroadpA}) to describe 
the transverse momentum broadening of partons and use the above parameters.

It is well-known that the PDFs are modified in the
nuclear environment. This is taken into account
by various shadowing parameterizations~\cite{Shadxnw_uj,EKS,HKM,Fran02}.
In the present work, we display results obtained with the 
EKS parameterization, which has an antishadowing
feature~\cite{EKS}, and with 
the updated HIJING parameterization~\cite{Shadxnw_uj}, which
incorporates different quark and gluon shadowing, and 
has an impact-parameter dependent and an 
impact-parameter independent version.
The impact-parameter dependence is taken into account
by a term $\propto (1-b^2/R_A^2)$, which re-weighs the shadowing
effect inside the nucleus.

The impact parameter dependence of the shadowing function
can influence $J/\psi$ production in $dAu$ collisions in
different rapidity windows~\cite{VogtPRL}. The investigation
of the centrality dependence of the pion production could
contribute  to the study of this interesting question.

\section{ Results on pion production in $dAu$ collisions}

Including the multiscattering and shadowing effects summarized
in the previous section one can
calculate the invariant cross section for pion production
in $dAu$ collision.
Moreover, introducing the nuclear modification factor
$R_{dAu}$, as
\be
\label{rdau}
R_{dAu} = \f{E_{\pi}\dd \sigma_{\pi}^{dAu}/\dd ^3p}
            {N_{bin} \cdot E_{\pi}\dd \sigma_{\pi}^{pp}/\dd ^3p} 
= \f{E_{\pi}\dd \sigma_{\pi}^{dAu}({\tt \ with \ nuclear \ effects})/\dd ^3p}
{E_{\pi}\dd \sigma_{\pi}^{dAu}({\tt \ no \ nuclear \ effects})/\dd ^3p} \,\, ,
\ee
nuclear effects can be investigated clearly and efficiently, using 
a linear scale. The value of $N_{bin}$ can be determined using the Glauber
geometrical overlap integral as in eq. (\ref{dAuX}). Here we apply
the right-most equation, which does not  require the determination of
$N_{bin}$ from the Glauber model.

\begin{figure}
\hspace{2.5truecm}
\resizebox{0.66\columnwidth}{!}{  
\includegraphics{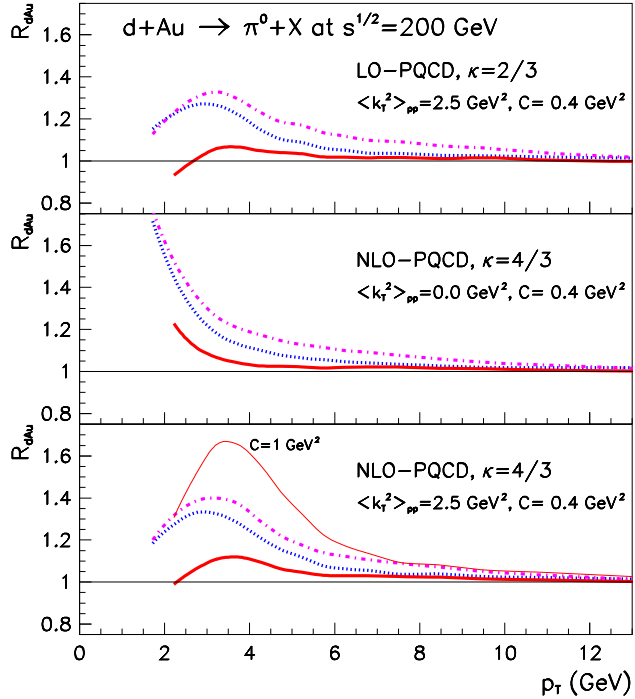}    
}   
\caption{ The $R_{dAu}$ nuclear modification factor 
for minimum bias $dAu$ collision in LO (upper panel),
in NLO without (middle panel) and with 
 (lower panel) intrinsic $k_T$.
For details see text and Fig. 1.
The dotted lines indicate the enhancement in $R_{dAu}$
connected to multiscattering only, which is
characterized by the average broadening per $NN$ collision, $C=0.4$ GeV$^2$/c$^2$.
Thick solid lines show the results after additionally including
a $b$-independent shadowing parameterization from the 
HIJING model~\cite{Shadxnw_uj}.
The thin line in the bottom panel was obtained with $C=1$ GeV$^2$/c$^2$.
The dash-dotted lines display the influence of 
the EKS-shadowing~\cite{EKS},
which has an anti-shadowing contribution, increasing $R_{dAu}$.}
\label{fig:2}       
\end{figure}

Figure 2 summarizes our results for the nuclear modification factor $R_{dAu}$ 
in minimum bias $dAu$ collisions at $\sqrt{s}=200$ AGeV.
In the top panel the LO case with finite
intrinsic $k_T$ is displayed. The center panel shows the NLO case
without intrinsic $k_T$ in the $pp$ collision,
corresponding to a well-focused initial beam of partons, 
but including the nuclear
broadening effect connected to multiscattering. The bottom panel shows the
NLO results with initial intrinsic $k_T$ in the $pp$ collision and
nuclear broadening.

The dashed lines indicate the Cronin enhancement connected to
the nuclear multiscattering characterized by $C=0.4$ GeV$^2$/c$^2$.
The peak structure with an enhancement of 25\%
is clearly seen in the case of large initial intrinsic $k_T$
(top and bottom panels), but in the middle panel the
Cronin peak is shifted to too small values of $p_T$, out of the range
of our pQCD calculations (this property was discussed 
in Ref.~\cite{bgg}).

The influence of shadowing is indicated by thick solid lines in Fig. 2: 
a relative decrease of $R_{dAu}$  
is obtained using a $b$-independent parameterization of 
shadowing applied in the HIJING model~\cite{Shadxnw_uj}.  
Altogether, an upto 10\% enhancement can be seen in the
3 GeV/c $< p_T <$ 5 GeV/c region, which disappears at higher $p_T$.
The height of the peak depends on the value of the 
multiscattering parameter $C$ at a fixed shadowing parameterization.
To illustrate this, we included a calculation with $C=1$ GeV$^2$/c$^2$
in the bottom panel of Fig.~2 (thin line). 
The position of the peak is related to the intrinsic $k_T$ value
in the $pp$ collision: smaller value leads to a shift to
smaller $p_T$~\cite{bgg}. The EKS parameterization~\cite{EKS}, 
which contains anti-shadowing,
leads to a small surplus in $R_{dAu}$, as expected (dash-dotted lines). 

\begin{figure}
\hspace{2.5truecm}
\resizebox{0.66\columnwidth}{!}{  
\includegraphics{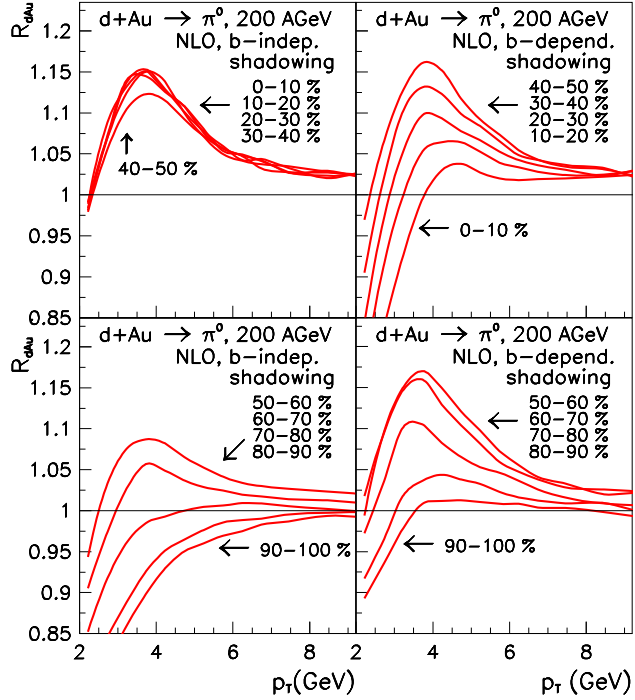}    
}   
\caption{ 
      The impact parameter dependence of the
nuclear modification factor $R_{dAu}$ for $b$-independent
shadowing (right column) and $b$-dependent shadowing (left column).
The lines correspond to different centrality
bins with an increment of 10\%. The 5 ``central'' curves 
(0-50\%) are shown in the upper panels, and the remaining 5 (``peripheral'') 
cases (50-100\%) can be seen in the lower panels.
}
\label{fig:3}       
\end{figure}

Figure 3 displays the centrality dependence of our NLO results 
($\kappa=4/3$, $\k2av_{pp}=2.5$ GeV$^2$/c$^2$, $C=0.4$ GeV$^2$/c$^2$)
for the nuclear modification factor $R_{dAu}$  
with the HIJING shadowing~\cite{Shadxnw_uj}. 
The left column shows the $b$-independent cases. Here,
the central bins (upper panel) yield essentially overlapping results,
because both, multiscattering and shadowing depend on the length
of the target matter (approximately constant in these
cases). Shadowing finally wins
moving toward peripheral collisions (lower panel),
because the Cronin effect disappears with decreasing $\nu_A(b)$,
but a slight shadowing is generated even in the most peripheral cases in
the recent parameterization of Ref.~\cite{Shadxnw_uj}.
However, there is a strong suppression in central collisions
in the $b$-dependent parameterization, 
which is not balanced by the Cronin effect (see upper right panel).
Shadowing is rapidly decreasing in this parameterization 
at larger impact parameter (bottom right panel),
and the Cronin effect becomes dominant for a certain $b$-range, 
increasing the nuclear modification factor. Finally, in peripheral
collisions the Cronin effect vanishes, and in lack of
strong shadowing the nuclear modification factor recovers to
unity at a reasonable transverse momentum value.
Precise data on the centrality dependence
in the nuclear modification factor could yield information about the
interplay between multiscattering (including saturated vs. non-saturated
parameterizations of multiscattering) and shadowing in the $dAu$ collision.

\begin{figure}
\hspace{2.5truecm}
\resizebox{0.66\columnwidth}{!}{  
\includegraphics{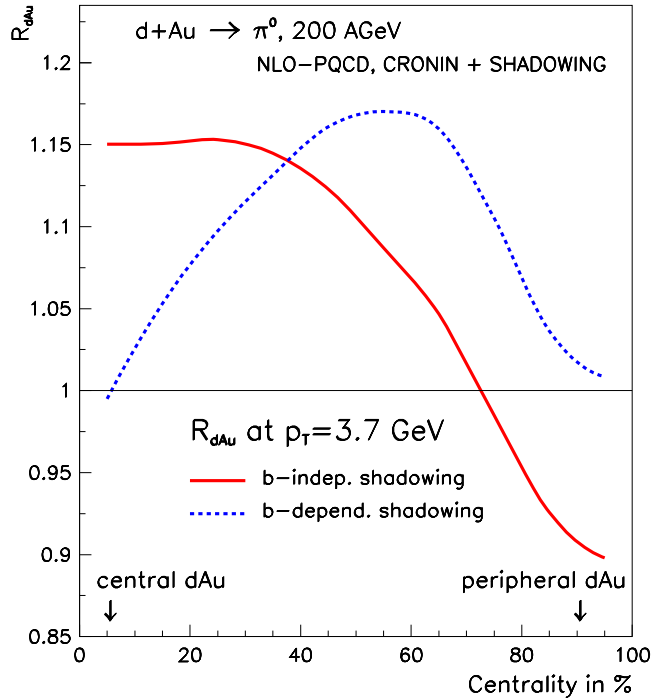}    
}   
\caption{ 
      The centrality dependence
of the nuclear modification factor
$R_{dAu}$ in the peak region, at fix $p_T=3.7$ GeV/c.
The solid line corresponds to the case of $b$-independent shadowing,
the dashed line describes $b$-dependent shadowing. 
}
\label{fig:4}       
\end{figure}

Figure 4 is constructed to emphasize the structure and properties of Fig.~3.
The centrality dependence of $R_{dAu}$ is shown at fix 
$p_T=3.7$ GeV/c, in the peak region of $R_{dAu}$. It is easy to appreciate 
the difference between the $b$-independent (solid line)
and the $b$-dependent shadowing (dashed line) from this Figure.

\begin{figure}
\hspace{2.5truecm}
\resizebox{0.66\columnwidth}{!}{  
\includegraphics{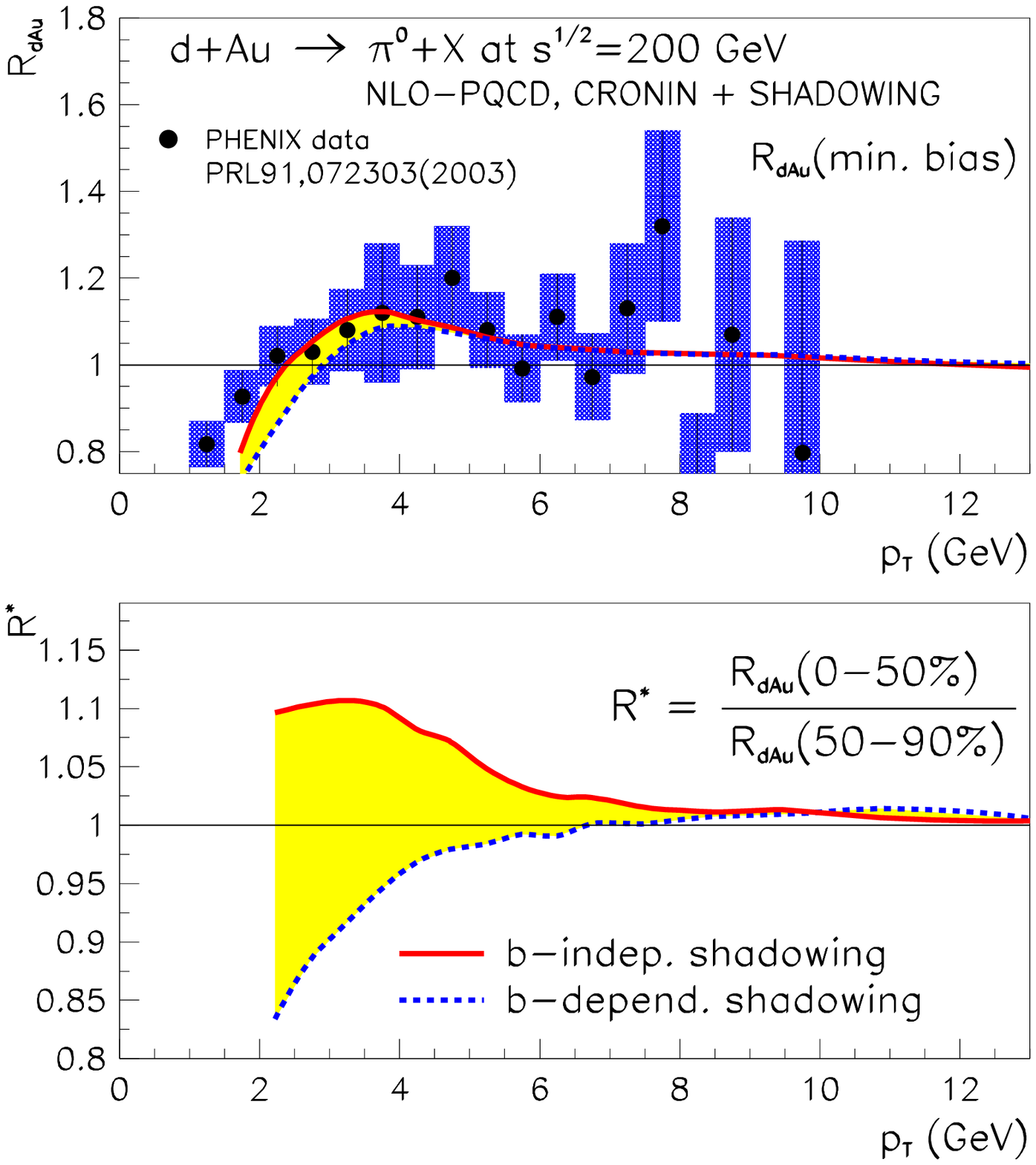}    
}   
\caption{ 
      Top panel: the nuclear modification factor
$R_{dAu}$ for minimum bias $dAu$ collisions obtained from
our NLO calculation in case of $b$-independent (solid line)
and $b$-dependent shadowing (dashed line). 
Experimental data are from the 
PHENIX Collaboration~\cite{RHIC_user03_PHENIX}.
Bottom panel: the double ratio of the nuclear modification
factors for central to peripheral collision, 
$R^* = R_{dAu}(0-50 \%) / R_{dAu}(50-90 \%)$.
The full line indicates the ratio $R^*$
for the $b$-independent shadowing, the dashed line corresponds to
the $b$-dependent shadowing.  
}
\label{fig:5}       
\end{figure}

Figure 5 (top panel) displays our results for the nuclear modification factor
in minimum bias $dAu$ collision for $b$-independent and $b$-dependent
shadowing parameterizations from HIJING.
Here we have $\kappa=4/3$ for the scales in the NLO calculation, 
and we use $\langle k_T^2 \rangle_{pp} = 2.5$ GeV$^2$/c$^2$ and $C=0.4$ GeV$^2$/c$^2$,
as in the bottom panel of Fig.~2.
In spite of the very different impact parameter dependence
of $R_{dAu}$ the minimum bias results
are very close to each other. This is because 
many details are averaged out in minimum bias
data, and the final result is no longer sensitive to the
$b$-dependence of shadowing. These results can be directly compared 
to the minimum bias experimental data.
In $h^{\pm}$ minimum bias data we expect a larger 
enhancement in the nuclear modification factor at this energy,
due to the presence of protons and antiprotons, which are known
to yield anomalous $p/\pi$ ratios~\cite{anom}. Uncertainties in the
treatment of fragmentation into protons~\cite{zlf02} preclude a
detailed prediction of $R_{dAu}$ for $h^{\pm}$ at the present time. 
On the other hand, parton fragmentation may overlap with parton
coalescence in the intermediate transverse momentum range, 
as it was investigated in $AuAu$ collisions~\cite{RCHwa,RCLevai,RCMuller},
and this effect may have influence in $dAu$ collisions.

In the bottom panel of Figure 5
we display the double ratio of the nuclear modification
factors for central to peripheral collision, 
$R^* = R_{dAu}(0-50 \%) / R_{dAu}(50-90 \%)$.
The full line indicates the ratio $R^*$
for the $b$-independent shadowing and the dashed line corresponds to
the $b$-dependent shadowing. In the window 
2  GeV/c $< p_T <$ 3 GeV/c the obtained difference is $\sim 30 \%$,
which could be seen in case of high precision data are available.

\begin{figure}
\hspace{0.9truecm}
\resizebox{0.93\columnwidth}{!}{  
\includegraphics{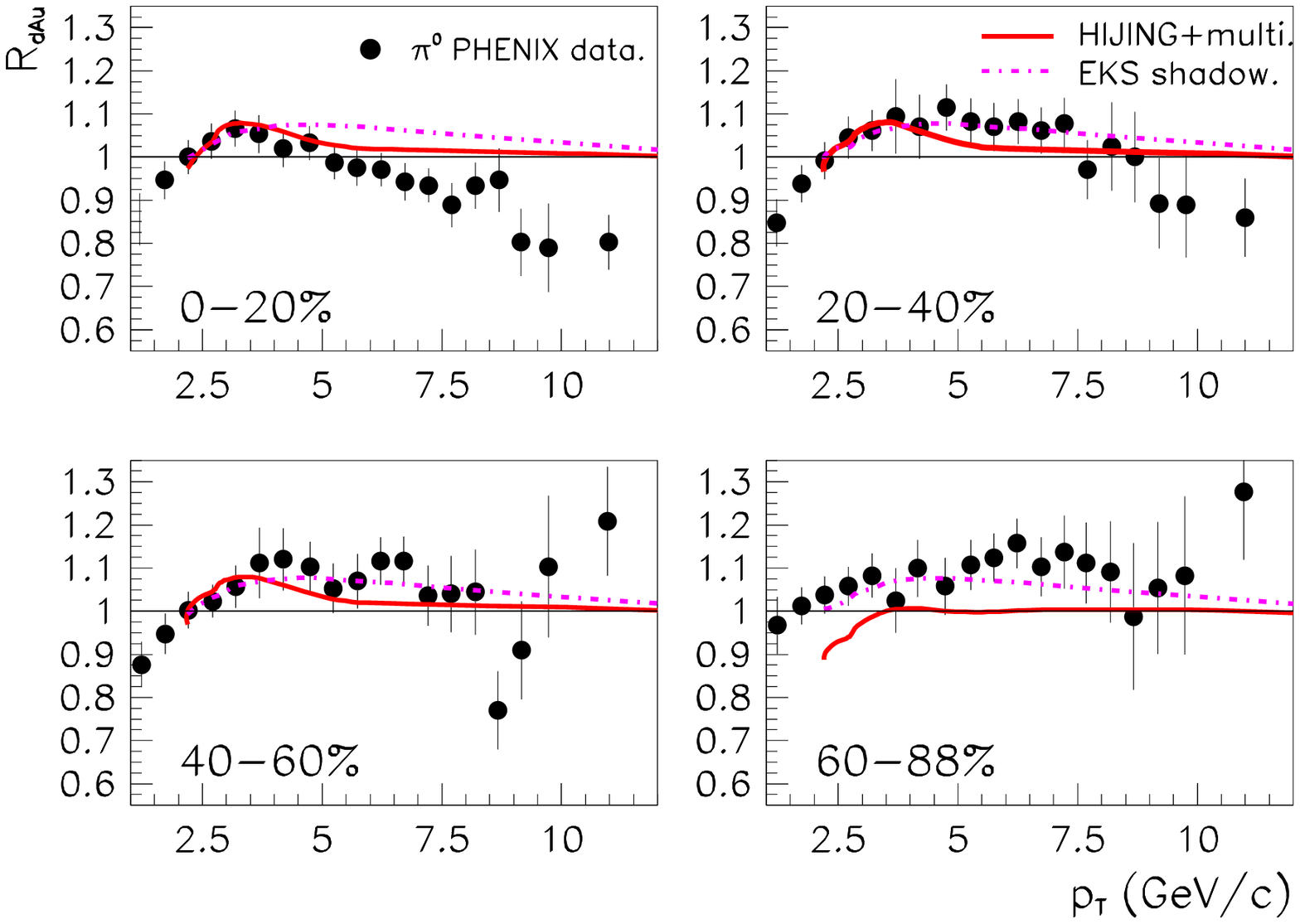}    
}   
\caption{ 
       The nuclear modification factor
$R_{dAu}$ for pion production in $dAu$ collisions 
at different centralities measured by 
PHENIX Collaboration~\cite{PHENIXdAu07centr}.
Results from our NLO calculation are displayed
with $b$-independent HIJING shadowing (solid line).
Dashed line corresponds to $b$-independent EKS shadowing.  
}
\label{fig:6}       
\end{figure}

Figure 6 displays recent high precision experimental data 
at different centralities on pion production
in $dAu$ collisions at $\sqrt{s_{NN}}=200$ GeV~\cite{PHENIXdAu07centr}.
We applied our model and calculated pion production, including
HIJING shadowing (solid lines) and EKS shadowing (dash-dotted lines).
Although the size of the error bars does not allow us to draw a 
firm conclusion about centrality dependence, selecting
transverse momentum windows at $p_T \approx 3.7$ GeV/c, we may see
a tendency similar to the dashed line in Fig. 4.
However, in a recent paper~\cite{EMC07},
applying the HKM shadowing function, we investigated in detail
the most central $dAu$ data and have found an indication of the
presence of jet energy loss in cold matter.
This finding means that theoretical descriptions have to
deal with nuclear shadowing, multiscattering, and jet energy loss 
already in dAu collisions, and may possibly lead to a structure 
similar to the full line in Fig. 4.
This interplay can be investigated quantitatively
if the precision of the data will become even higher by the analysis 
of a forthcoming $dAu$ run.

\section{ Discussion}

Summarizing our theoretical results in a NLO pQCD-based parton model
calculation for the $dAu$ collision, we see a clear
enhancement ($\approx 10-15$\%) in the 
nuclear modification
factor $R_{dAu}$ after including nuclear multiscattering and 
HIJING shadowing.
Our NLO calculation of pion production in the $dAu$ collision confirms
earlier LO results, while providing
an updated description of the invariant cross section and 
the interplay between multiscattering and shadowing effects.
We have found that there is a qualitative difference between 
position-independent and position-dependent shadowing; 
the specific structure of
the centrality dependence of the nuclear modification
factor results in maximum enhancement in semi-central collisions for 
position-dependent shadowing.
Thus, the centrality dependence of $R_{dAu}$ may be used
to obtain detailed information about 
the effects of multiscattering and shadowing.
 
Recent experimental data on $\pi^0$ production~\cite{PHENIXdAu07centr}
allow us to perform a detailed investigation of
centrality dependence. Although the minimum bias 
$\pi^0$ data display a 10-15\% increase in the
nuclear modification factor supporting the idea of interplay
between multiscattering and shadowing in the reproduction of the
Cronin effect, but detailed conclusion can not be obtained
at the recent precision of the data.
The $h^{\pm}$ data~\cite{PHENIXdAu06,PHENIXdAu07} appear to have an even 
larger increase, but we do not have a theory comparison 
for these data at present, because the proton (and antiproton) 
production is not well described by recent pQCD based parton models.

\bigskip

Acknowledgments:
We thank G. David, M. Gyulassy, I. Vitev, and X.N. Wang 
for useful comments and stimulating discussions.  We are
grateful to P. Aurenche and his collaborators for their NLO code,
which served as the basis of calculational developments reported here.
This work was supported in part by  U.S. DOE grants DE-FG02-86ER40251,
DE-FG02-93ER40764, MTA-OTKA-NSF grant INT-0000211 and 
Hungarian grants T034842, T043455, NK062044, IN71374.
Supercomputer time provided by BCPL in Norway and the EC -- Access to
Research Infrastructure action of the Improving Human Potential programme
is gratefully acknowledged.


\begin{thebibliography}{99}
\bibitem{note}
In the NLO calculation displayed in Fig. 2c of 
Ref.~\cite{PHENpi0200pp}, 
the scale choice $Q=Q_R=Q_F=p_T$
appears to be a good general fit within error bars; however, concentrating
on the central values at high $p_T$ would prefer scales like 
$Q=Q_R=Q_F=(5/3)p_T$, leading to a factor 2 underestimation of the data
at low $p_T$, just like in our calculation.
\end{thebibliography}
\end{document}